\documentclass[manuscript]{acmart}

\usepackage[dvipsnames]{xcolor}
\usepackage{array}
\usepackage{graphicx}
\usepackage[table,xcdraw]{xcolor}
\usepackage{colortbl}
\usepackage{longtable}
\usepackage{subcaption}
\usepackage{float}
\usepackage{longtable}
\usepackage{booktabs}
\usepackage{xcolor}
\usepackage{tcolorbox}
\usepackage{listings}
\usepackage{multirow}
\usepackage{makecell}
\usepackage{xspace}
\AtBeginDocument{%
  }

\newcommand{\gtalong}{Graduate TA\xspace}

\newcommand{\utalong}{Undergraduate TA\xspace}

\newcommand{\solution}{\textbf{Sample Solution}\xspace}

\newcommand{\naiveX}{\textbf{Naïve (Naï)}\xspace}
\newcommand{\fewshotX}{\textbf{Few-shot (FSh)}\xspace}
\newcommand{\solutionX}{\textbf{Sample Solution (Sol)}\xspace}
\newcommand{\chainotX}{\textbf{Chain of Thought (CoT)}\xspace}

\newcommand{\othreemini}{\texttt{o3-mini}\xspace}

\setcopyright{none}

\newcommand{\xhdr}[1]{\vspace{1.5mm}\noindent\textbf{#1}}

\begin{document}


\title{When LLMs Help -- and Hurt -- Teaching Assistants in Proof-Based Courses}
%

\author{Romina Mahinpei}
\email{rmahinpei@princeton.edu}
\affiliation{%
  \institution{Princeton University}
  \city{Princeton}
  \state{New Jersey}
  \country{USA}
}

\author{Sofiia Druchyna}
\email{sd0937@princeton.edu}
\affiliation{%
  \institution{Princeton University}
  \city{Princeton}
  \state{New Jersey}
  \country{USA}
}

\author{Manoel Horta Ribeiro}
\email{manoel@cs.princeton.edu}
\affiliation{%
  \institution{Princeton University}
  \city{Princeton}
  \state{New Jersey}
  \country{USA}
}

\renewcommand{\shortauthors}{Mahinpei et al.}

\begin{abstract}
\noindent 

\end{abstract}





\begin{abstract}
Teaching assistants (TAs) are essential to grading and feedback provision in proof-based courses, yet these tasks are time-intensive and difficult to scale. Although Large Language Models (LLMs) have been studied for grading and feedback, their effectiveness in proof-based courses is still unknown. Before designing LLM-based systems for this context, a necessary prerequisite is to understand whether LLMs can meaningfully assist TAs with grading and feedback. As such, we present a multi-part case study functioning as a technology probe in an undergraduate proof-based course. We compare rubric-based grading decisions made by an LLM and TAs with varying levels of expertise and examine TAs’ perceptions of feedback generated by an LLM. We find substantial disagreement between LLMs and TAs on grading decisions but that LLM-generated feedback can still be useful to TAs for submissions with major errors. We conclude by discussing design implications for human-AI grading and feedback systems in proof-based courses.
\end{abstract}

\begin{teaserfigure}
    \centering
    \includegraphics[width=0.8\linewidth]{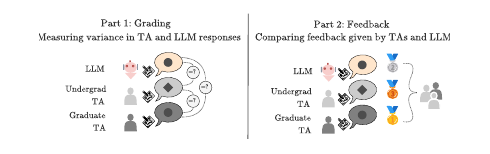}
    \Description{}
    \caption{\textbf{We explore whether Large Language Models (LLMs) can assist teaching assistants through a multi-part case study in a proof-based course.} In Part 1, we compare the grading of our LLM pipeline and TAs of varying expertise levels, assessing where disagreements emerge.
    In Part 2, we collect TA preference rankings for feedback provided by other TAs of varying levels of expertise as well as feedback provided by our LLM pipeline.}
    \label{fig:studies_overview}
\end{teaserfigure}

\maketitle

\section{Introduction}
Teaching assistants (TAs) play a fundamental role in the university system, particularly in grading assignments and providing feedback to students~\cite{Ahmed2024}. In proof-based courses, these responsibilities are particularly demanding given the struggles students face with the course material ~\cite{Stylianou2015,Lew2019}. At scale, these challenges place significant time and cognitive burdens on TAs and can lead to inconsistencies across graders~\cite{Messer2024,Doe2013}. At the same time, large language models (LLMs) have attracted interest in educational contexts. Prior work has shown that LLMs can approximate human grading in programming and short-answer tasks and can generate feedback that students perceive as comparable to instructor feedback~\cite{Floden, Paraskevas, Chamuditha, Qiu, Lee, Meyer,Mello25}. However, these findings primarily come from domains with relatively constrained solution spaces or clear correctness criteria. Whether and how LLM-backed systems can support proof-based courses, where solutions are open-ended, sequential, and sensitive to course-specific conventions, remains largely unexplored.

Rather than positioning LLMs as replacements for humans, we envision their role as \textit{teaching assistant companions} that support existing TA workflows in proof-based courses. Before such systems can be responsibly designed, it is necessary to understand where LLMs align with or diverge from human grading judgment and how TAs perceive the usefulness of LLM-generated feedback in practice. To this end, we present a multi-part case study functioning as a technology probe in an undergraduate proof-based course that examines (1) how LLM rubric-based grading compares to that of graduate and undergraduate TAs and (2) how TAs evaluate LLM-generated feedback on student proofs.

Our findings point to a clear asymmetry between grading and feedback that has direct implications for the design of LLM-based tools to support TA workflows. While our LLM pipeline (see section \ref{sec:llm_pipeline}) frequently diverges with human graders in rubric item assignment, TAs often find the feedback generated by our LLM pipeline to be helpful in cases involving major errors. At the same time, TAs express concerns about verbosity when feedback is generated for largely correct solutions. Together, these observations suggest that LLM-powered systems for proof-based courses should not treat grading and feedback as a single, unified task. Instead, they motivate designs that separate evaluative judgment from formative support, positioning LLMs as TA companions rather than replacements.

\vspace{-0.25cm}
\section{Related Works}
Recent work has explored the use of LLMs for grading across a variety of educational contexts. Prior studies show that LLMs can generate plausible grades and exhibit human-like variability when evaluating university-level exams, programming assignments, and short-answer tasks ~\cite{Floden, Paraskevas, Mello25}. However, these systems often struggle with alignment to human graders, including inconsistent rubric interpretation, over- or under-penalization of errors, and sensitivity to prompt design ~\cite{Mello25}. To mitigate these issues, researchers have explored rubric-guided grading, retrieval-augmented generation, and structured prompting strategies, finding improvements in accuracy and consistency in constrained domains such as programming and science education ~\cite{Chamuditha, Qiu, Lee}.

Beyond grading, LLMs have also been investigated as feedback providers. In programming education, LLMs demonstrate strong error detection and explanation capabilities, though aligning their feedback with instructional goals remains challenging ~\cite{Paraskevas}. Furthermore, studies of student perceptions suggest that students often perceive AI-generated feedback as comparable to instructor feedback, unless it becomes repetitive or overly verbose [13, 18, 22]. Other work highlights the potential of LLMs to support human feedback workflows at scale, particularly when used as a supplement rather than a replacement for human judgment ~\cite{Sylvio25, Shin25}.

Despite this growing body of work, proof-based courses in higher education remains largely unexplored. Grading mathematical proofs requires evaluating open-ended, sequential reasoning and applying course-specific conventions that are often implicit rather than formally specified. Moreover, existing studies rarely account for variability among human graders based on their level of expertise. We address these gaps by presenting a case study in a proof-based undergraduate course that examines both rubric-based grading agreement across graduate and undergraduate TAs and TA perceptions of LLM-generated feedback and those produced by graduate and undergraduate TAs.

\vspace{-0.25cm}
\section{Methods}
\begin{figure}[t]
    \centering
    \includegraphics[width=0.6\textwidth]{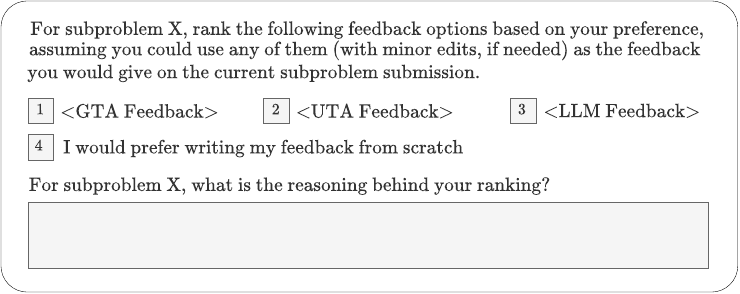}
        \caption{\textbf{Sample survey question} asking participants to rank the available feedback options for a student's submission to a subproblem.}
    \label{fig:survey_question} 
\end{figure}

\begin{figure}[t]
    \centering
    \includegraphics[width=0.7\textwidth]{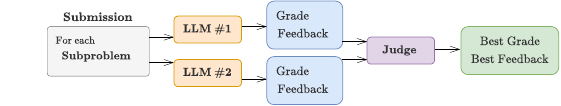}
    \caption{\textbf{LLM-as-a-Judge architecture for rubric-based grading and feedback provision.} Two \othreemini instances independently generate candidate outputs for each subproblem while a third \othreemini  instance selects the best response.}
    \label{fig:architecture} 
\end{figure}

\subsection{Study Design}
We conduct a multi-part case study in Course X, an undergraduate proof-based course in computer science on economics at computing at a private U.S. institution. Assessments in Course X include \textbf{long-answer proof problems} that require students to construct open-ended solutions and are graded using rubric-based criteria for \textbf{solution quality (SQ)} and \textbf{writing quality (WQ)}. Each rubric consists of multiple disjoint items reflecting varying degrees of correctness and clarity (see Appendix \ref{app:sample_materials} for a sample long-answer problem and its rubrics). To evaluate the potential role of LLMs in TA workflows, we select the most challenging proof question from a mid-semester assignment of Course X, consisting of five sequential subproblems. We randomly sample 30 student submissions and recruit three graduate TAs (GTAs) and three undergraduate TAs (UTAs) from Course X. We then independently grade and annotate each submission with feedback by one GTA, one UTA, and our LLM pipeline (see section 3.2).

With data collection complete, our study proceeds with two follow-up parts. \textbf{Part 1 (Grading)} examines agreement between the GTAs, UTAs, and our LLM pipeline from the previous step in rubric item assignment across all subproblems and rubric categories. \textbf{Part 2 (Feedback)} evaluates TA perceptions of the LLM-generated feedback and those produced by GTAs and UTAs from the previous step. We recruit 13 additional TAs who had previously served in Course X to participate in a survey comparing feedback written by the GTAs, UTAs, and our LLM pipeline, as well as the option to write feedback from scratch (see Figure \ref{fig:survey_question}). For each subproblem, participants rank feedback options and provide brief rationales. Feedback sources are anonymized and presented in randomized order to mitigate bias.

\subsection{LLM Pipeline}
\label{sec:llm_pipeline}
Our LLM pipeline uses OpenAI's \othreemini model and grading and feedback prompts (see Appendix \ref{app:final_prompt} for prompts) that provide instructions on the respective tasks along with the problem statement, grading rubrics, instructor solution, and student answer. We select our model and refine our prompts through a series of benchmarking studies (see Appendix \ref{app:benchmarking} for benchmarking) done with Course X in preparation for the case study. To improve robustness, we also adopt an LLM-as-a-Judge setup (see Figure \ref{fig:architecture}): two \othreemini instances independently select rubric items and generate feedback for each subproblem, and a third \othreemini instance selects the best response for grading and for feedback. 
The selected outputs are treated as final, and each LLM has access to earlier subproblem contexts within the same submission. 

\vspace{-0.25cm}
\section{Results}
\subsection{Part 1: Grading}
\begin{table}[t]
\footnotesize
    \centering
    \caption{Percentage agreement between all possible pairwise groupings of \gtalong, \utalong, and LLM across all five subproblems for our two rubric categories along with the percentage agreement between all possible pairwise groupings after pooling all subproblem submissions into a single set. Each metric includes a 95\% confidence interval (shown in brackets).}
    \begin{tabular}{|c|c|c|c|c|} 
    \hline
    \textbf{Rubric Type} & \textbf{Subproblem} & \makecell{\textbf{GTA-UTA} \\ \textbf{Agreement \%}} & \makecell{\textbf{GTA-LLM} \\ \textbf{Agreement \%}} & \makecell{\textbf{UTA-LLM} \\ \textbf{Agreement \%}} \\
    \hline 
    \multirow{6}{*}{\makecell{\textbf{Solution Quality} \\ \textbf{(SQ)}}} & \textbf{A} & 96.67\% [87, 100]& 100.00\% [100, 100] & 96.67\% [87, 100] \\
    & \textbf{B} & 100.00\% [100, 100] & 73.33\% [60, 88] & 73.33\% [57, 90] \\
    & \textbf{C} & 73.33\% [53, 87] & 63.33\% [47, 80] & 63.33\% [47, 78] \\
    & \textbf{D} & 70.00\% [53, 87] & 66.67\% [52, 82] & 46.67\% [32, 60] \\
    & \textbf{E} & 93.33\% [83, 100] & 33.33\% [20, 47] & 33.33\% [18, 50] \\
    \hline 
    & \textbf{Pooled} & \textbf{86.67\%} \textbf{[77, 95] } & \textbf{67.33\%} \textbf{[47, 83]}  & \textbf{62.67\%} \textbf{[42, 78]}  \\
    \hline
    \multirow{6}{*}{\makecell{\textbf{Writing Quality} \\ \textbf{(WQ)}}} & \textbf{A} & 100.00\% [100, 100]& 100.00\% [100, 100] & 100.00\% [100, 100] \\
    & \textbf{B} & 93.33\% [87, 100] & 80.00\% [67, 93] & 73.33\% [57, 87] \\
    & \textbf{C} & 93.33\% [83, 100] & 80.00\% [63, 93] & 76.67\% [63, 92] \\
    & \textbf{D} & 96.67\% [90, 100] & 86.67\% [75, 97] & 83.33\% [72, 93] \\
    & \textbf{E} & 93.33\% [82, 100] & 63.33\% [50, 80] & 60.00\% [47, 78] \\
    \hline 
    & \textbf{Pooled} & \textbf{95.33\%} \textbf{[87, 100]} & \textbf{82.00\%} \textbf{[70, 93]} & \textbf{78.67\%} \textbf{[63, 90]}  \\
    \hline
    \end{tabular}
    \label{tab:study2_combined}
\end{table}

We provide our results for both rubric categories across all subproblems in Table \ref{tab:study2_combined}. These results show that in most cases, all pairwise groupings experience some degree of disagreement. 
The GTA-UTA group has the highest agreement across all sub-problems and when pooled for both rubric categories ({SQ}:~86.67\%; {WQ}:~95\%), 
followed by GTA-LLM ({SQ}:~67.33\%; {WQ}:~82\%), 
and UTA-LLM ({SQ}:~62.67\%; {WQ}:~78.67\%). We also observe a large drop in agreement between the LLM and GTAs as well LLMs as UTAs for later dependent subproblems [\textbf{O1A: Agreement Drops in Later Subproblems}]. For instance, while agreement for solution quality between GTAs and UTAs for subproblem E is 93.33\%, it is 33.3\% for GTAs and LLM (same for UTAs and LLM).

Looking at all submissions where grading divergences happen, we find that disagreements between GTAs and UTAs often occur when either of the graders miss an error in the submission. 
On the contrary, in disagreements between LLM and GTAs as well as LLM and UTAs, the LLM would often correctly identify the presence of mistakes but grade them more harshly by assigning a lower rubric item, likely due to the fact that rubric assignment has some subjectivity in deciding whether an error is `major` or `minor' [\textbf{O1B: LLM's Stricter Rubric Interpretation}].
In addition, the LLM often requires a student's solution to be more thorough and explicit [\textbf{O1C: LLM's Expectation of Explicit Reasoning}]. For example, if a student skips a few logical steps in their proof, the LLM would assign a lower rubric item even if the logical skips align with conventions taught in the course.
Lastly, we observe that the LLM sometimes directly compares the student's submission to the provided instructor's solution and assigns lower rubric items if the student's approach differs significantly [\textbf{O1D: LLM's Sensitivity to Solution Strategy}].

\vspace{-0.25cm}
\subsection{Part 2: Feedback}
We provide the tallies for the different feedback options in Table \ref{tab:study3_counts}, separating scenarios where different feedback options are present (e.g., no GTA nor UTA feedback). This variation arises because TAs sometimes choose not to provide feedback, resulting in some submissions where not all options were present.

When neither GTA nor UTA feedback is provided (block one), participants slightly prefer writing feedback from scratch over LLM-generated feedback (60 out of 112 rankings). 
In contrast, when at least one human grader provides feedback (blocks two to four), this pattern reverses: LLM feedback is preferred on par with UTA feedback and ranked higher than feedback from scratch [\textbf{O2A: Preference for LLM Feedback in the Presence of Human Feedback}]. Specifically, when all human graders provide feedback (block four) the LLM feedback option is most frequently ranked first (28 out of 81).

Our qualitative analysis of the participants' rationales helps explain these patterns. 
Participants frequently describe LLM feedback as a useful starting point for addressing issues in student work [\textbf{O2B: LLM Feedback as a Starting Point}]. One participant notes, \textit{“I think the LLM feedback is very rigorous in terms of what can be improved, and explaining what is wrong,”} while another shared, “\textit{This is very detailed feedback, especially for the structure of proof. Helpful in the context of making them better proof writers.”} However, when submissions contain no major errors, participants often feel that LLM feedback lacks conciseness and included unnecessary detail [\textbf{O2C: Verbose LLM Feedback for Correct Submissions}]. As one participant explains, \textit{“Since the solution is correct, I would write a shorter response.”} Another disagrees with the LLM’s suggestion to add additional explanation, stating, \textit{“Don’t think additional explanations are required. The steps are pretty clear.”}
This explains why LLM feedback is not preferred over feedback from scratch when no human feedback is available: LLMs provide verbose feedback when little or none is needed.

\begin{table}[t]
\small
    \caption{Tallies of the preference rankings for the different feedback options. The first block provides the counts for subproblem submissions that received no GTA nor UTA feedback, the second block provides the counts for subproblem submissions that received GTA feedback but no UTA feedback, the third block provides the counts for subproblem submissions that received TA feedback but no GTA feedback, and the fourth block provides the counts for subproblem submissions that received both GTA and UTA feedback.}
    \label{tab:study3_counts}
    \begin{tabular}{|c|c|c|c|c|} 
    \hline
    \textbf{Feedback Option} & \textbf{Rank 1 Count} & \textbf{Rank 2 Count} & 
    \textbf{Rank 3 Count} & 
    \textbf{Rank 4 Count} \\
    \hline 
    \textbf{LLM} & 52 & 60 & --- & --- \\
    \textbf{Scratch} & 60 & 52 & --- & --- \\
    \hline 
    \textbf{LLM} & 10 & 5 & 16 & ---  \\
    \textbf{GTA} & 13 & 13 & 5 & ---  \\
    \textbf{Scratch} & 8 & 13 & 10 & ---  \\
    \hline
    \textbf{LLM} & 33 & 29 & 39 & ---  \\
    \textbf{UTA} & 44 & 19 & 38 & ---  \\
    \textbf{Scratch} & 24 & 53 & 24 & --- \\
    \hline 
    \textbf{LLM} & 28 & 11 & 19 & 23 \\
    \textbf{GTA} & 25 & 20 & 19 & 17 \\
    \textbf{UTA} & 15 & 19 & 23 & 24 \\
    \textbf{Scratch} & 13 & 31 & 20 & 17 \\
    \hline
    \end{tabular}

\end{table}

\vspace{-0.25cm}
\section{Discussion}
The goal of our case study is to understand whether LLMs can meaningfully assist TAs in proof-based courses, serving as a technology probe for future human-AI systems targeting grading and feedback provision. Based on the observations surfaced in the Results [\textbf{O1A–O2C}], we argue that LLMs expose a fundamental asymmetry between \textit{evaluative} and \textit{formative} work in proof-based courses. While LLM behavior often clashes with TA judgment in grading, those same behaviors become assets when repurposed for feedback, provided appropriate human control and interaction design.

\xhdr{Grading as Situated Judgment, Not Rubric Execution}. Observations from Part 1 show that grading disagreements are not uniformly distributed but intensify for later, dependent subproblems [\textbf{O1A}]. This pattern highlights that grading proofs is inherently contextual: later judgments depend on how earlier errors are interpreted, forgiven, or propagated. Human graders (both GTAs and UTAs) often exercise discretion by accommodating skipped steps, interpreting student intent, and calibrating severity when mistakes compound across subproblems. In contrast, the LLM consistently applies stricter rubric interpretations [\textbf{O1B}], requires explicit reasoning even when omissions align with course conventions [\textbf{O1C}], and penalizes deviations from the instructor’s reference solution [\textbf{O1D}]. These behaviors suggest that the LLM treats grading as a \textit{static classification task} rather than a \textit{situated practice} by instructional conventions and pedagogical goals. These observations suggest caution in the deployment of LLMs as autonomous graders in proof-based courses. If incorporated into grading workflows, LLMs may be better suited to surface potential issues or inconsistencies for human review, rather than assigning final rubric categories.

\xhdr{Feedback as a Potential Site for Human-AI Collaboration}. Observations from Part 2 suggest that LLMs can meaningfully support feedback provision under specific conditions. When at least one human grader has already provided feedback, participants frequently prefer LLM-generated feedback over writing feedback from scratch [\textbf{O2A}]. Qualitative rationales indicate that TAs valued LLM feedback for its thoroughness and clarity, particularly for submissions with major errors [\textbf{O2B}]. However, this same thoroughness becomes a liability for largely correct solutions. In those cases, participants describe LLM feedback as unnecessarily verbose and misaligned with their pedagogical intent to exercise restraint [\textbf{O2C}]. These observations highlight how LLMs are better positioned as feedback drafting partners rather than standalone feedback providers. Systems should enable TAs to quickly shorten, modify, or discard LLM-generated feedback, allowing models to reduce repetitive labor while preserving human control over tone, length, and instructional intent.

\xhdr{Designing for Asymmetry: Separating Evaluative and Formative Tasks.} Taken together, these observations suggest that a single, unified LLM-powered system is unlikely to work across grading and feedback tasks. Instead, future systems should explicitly recognize the asymmetry between \textit{evaluative judgment} and \textit{formative guidance}. Grading demands sensitivity to context and course-specific conventions, areas where LLMs currently struggle. Feedback, by contrast, benefits from the LLM’s tendency toward completeness and explicitness, as long as humans retain final authority. Such systems should give TAs fine-grained control over feedback detail, allowing them to calibrate guidance based on the quality of student work. To better align system behavior with instructional practice, course-specific conventions, such as expectations around skipped steps, solution strategies, and error severity, should be made explicit and editable. Finally, interaction designs must support maintaining context across sequential subproblems, helping both humans and AI reason consistently as earlier decisions propagate forward.

\xhdr{Limitations \& Future Work.} Our study is limited in scope to a single proof-based undergraduate course and a single LLM. While this limits generalizability, it enables a fine-grained analysis of human–AI interaction in a setting that relies on course conventions and adaptive decision-making. We view this tradeoff as appropriate for a case study intended to surface design insights rather than benchmark model performance. As for future work, an important direction is the design and deployment of interactive systems that operationalize the design implications identified in this study. For example, TA-facing tools could allow TAs to signal when minimal feedback is desired, edit LLM-generated feedback templates, or adjust how strictly rubrics are interpreted. Moreover, studying how TAs accept, modify, or reject LLM contributions in authentic workflows would provide stronger evidence of both usability and pedagogical impact and clarify how LLMs can best assist, rather than replace, human judgment in higher education.

\newpage
\bibliographystyle{ACM-Reference-Format}
\bibliography{references}

\newpage
\appendix

\section{Sample Proof Problem \& Rubrics}
\label{app:sample_materials}
Below, we provide a sample \textbf{long-answer proof problem} from Course X along with its corresponding grading rubrics, including both \textbf{solution quality (SQ)} and \textbf{writing quality (WQ)} rubrics.

\subsection{Problem Statement}
\label{app:problem_statement}
You are trying to collect all $n$ distinct cards from your favorite trading card game. The only way to
obtain new cards is to purchase a sealed pack containing one uniformly random card. After you buy a pack,
you open it, observe the card inside, and add it to your collection. If you have at least one copy of each
of the $n$ cards, you stop. Otherwise, you purchase a new pack. Prove that the expected number of packs you purchase is $\Theta(n \log n)$.

\subsection{Solution Quality Rubric}
Below, we provide the solution quality (SQ) rubric created by the instructor of Course X for the problem in \ref{app:problem_statement}.

\begin{table}[h]
\small
\centering
\label{tab:solution-quality-rubric}
\begin{tabular}{p{0.14\linewidth} p{0.63\linewidth} p{0.13\linewidth}}
\toprule
\textbf{Rubric Item ID} & \textbf{Rubric Item Description} & \textbf{Points} \\
\midrule
5 &
The solution is fully correct. In particular, there is a concrete outline (perhaps implicit, not necessarily explicitly stated) that is correct, and each step of the outline is implemented correctly with an appropriate level of detail. The logical flow of the solution is fully correct and clearly explained. It is not necessary to verify all calculations, as long as it is clear that the logic is sound and the final answer is correct.
& 8.0 \\
\midrule
4 &
The solution clearly contains all of the main ideas, but there are some minor (but non-trivial) issues with the logical flow. For example, the final answer may be correct, but some justifications are flawed or incomplete. The missing components should be identifiable.
& 6.5 \\
\midrule
3 &
The solution makes significant and concrete partial progress, but does not fully solve the problem. Brief feedback should be provided to indicate how the existing work could be generalized or extended.
& 5.0 \\
\midrule
2 &
The solution makes some concrete partial progress, but is still far from a complete solution. Brief feedback should be provided to help guide further progress.
& 3.0 \\
\midrule
1 &
The solution does not make concrete partial progress toward a solution, but demonstrates a reasonable understanding of the problem being asked.
& 1.0 \\
\midrule
0 &
The solution is absent or fundamentally flawed.
& 0.0 \\
\bottomrule
\end{tabular}
\end{table}

\subsection{Writing Quality Rubric}
Below, we provide the writing quality (WQ) rubric created by the instructor of Course X for the problem in \ref{app:problem_statement}.

\begin{table}[h]
\small
\caption{Writing Quality (WQ) rubric for the long-answer proof problem provided in \ref{app:problem_statement}}
\centering
\label{tab:writing-quality-rubric}
\begin{tabular}{p{0.14\linewidth} p{0.63\linewidth} p{0.13\linewidth}}
\toprule
\textbf{Rubric Item ID} & \textbf{Rubric Item Description} & \textbf{Points} \\
\midrule
4 &
The solution is written clearly. The logical flow is easy to follow, and each step of the outline is implemented with appropriate detail. The outline need not be explicitly stated, as long as it is recoverable from the text. Minor grammatical errors, typos, or calculation mistakes are acceptable if they do not interfere with understanding the logical flow.
& 2.0 \\
\midrule
3 &
The solution is written fairly clearly. The overall outline is understandable, but some parts of the logical flow may be tricky to follow.
& 1.5 \\
\midrule
2 &
The solution is written adequately but requires some effort to understand the logical flow. Large portions may be challenging to interpret, even if some parts are clear.
& 1.0 \\
\midrule
1 &
Some parts of the solution are clear and easy to follow, but much of the solution requires significant effort to understand.
& 0.5 \\
\midrule
0 &
The solution is difficult to evaluate. Claims may be imprecise or incorrect, and it may be hard to determine the intended logical flow or the concrete progress made.
& 0.0 \\
\bottomrule
\end{tabular}
\end{table}

\newpage
\section{Prompts for LLM Pipeline}
\label{app:final_prompt}
As part of our LLM-pipeline, two independent instances of \othreemini are first asked to grade (using the prompt in \ref{app:grading_prompt}) and then provide feedback (using the prompt in \ref{app:feedback_prompt}) for a given student submission. A third \othreemini instance is then asked to act as a judge (using the prompt in \ref{app:judge_prompt}) and select the best grade and feedback.

\subsection{Grading Prompt}
\label{app:grading_prompt}
You are a grading assistant in a proof-based course on economics and computing. The user will provide you with a problem statement, the corresponding grading rubric, sample solution, and a student answer. Your job is to use these items to evaluate a student answer to the problem and assign a grade.

\smallskip
\noindent The grading rubric consists of multiple categories. The student answer should be graded based on all these categories.
For each category, there is a list of disjoint rubric items, out of which one should be selected (i.e., you should exactly select one rubric item per category). 
To get the student's final grade on the problem, add up the values in the pointDelta field from each of the rubric items you select. Note that the student's solution does NOT need to exactly match the sample solution. The sample solution is only meant to serve as a reference; it DOES NOT HAVE TO MATCH the submission you are grading.

\smallskip
\noindent Respond to the user only with a valid JSON object:
\begin{verbatim}
{
  "sid": "{student_id}",
  "grade": <numerical_grade>,
  "rationale": [chosen_rubric_item_id_1, chosen_rubric_item_id_2]
}
\end{verbatim}

\noindent \textbf{Problem Statement}: \texttt{\{problem\}}

\noindent \textbf{Grading Rubric}: \texttt{\{rubric\}}

\noindent \textbf{Instructor Solution}: \texttt{\{instructor\_soln\}}

\noindent \textbf{Student Answer}: \texttt{\{student\_answer\}}

\subsection{Feedback Prompt}
\label{app:feedback_prompt}
Please provide a few sentences of feedback based on the grade you gave on how the student can improve their solution.

\smallskip
\noindent You can use a variety of strategies such as: hints/explanations on technical terms, examples illustrating the concept, hints/explanations on the conceptual context, hints/explanations on concept attributes, attribute-isolation examples, indications of mistakes, location of mistakes, hints/explanations on types of errors, hints/explanations on sources of errors, guidance for correcting errors, hints/explanations on task-specific strategies, hints/explanations on task-processing steps,
guiding questions, worked-out examples.

\smallskip
\noindent Do not simply mention that something is incorrect—provide actionable and constructive feedback that guides the student in refining their approach. Tailor the feedback to their specific errors while encouraging deeper understanding.

\subsection{Judge Prompt}
\label{app:judge_prompt}
Please act as an impartial judge and evaluate the quality of the responses, each consisting of a chosen rubric and feedback, provided by 2 LLM assistants to the student answer for the given problem statement. Your evaluation should consider correctness and helpfulness of the rubric items and feedback.

\smallskip
\noindent Your job is to evaluate which LLM feedback is better. You should independently solve the proof question shown in problem statement, look at the grading rubric and the student answer, and the reference solution. Note that the student solution does not need to match the reference solution.
Then evaluate the 2 LLM responses that provide grading rationale and a feedback to the student.

\smallskip
\noindent The grading rubric consists of multiple categories. The student answer is graded based on all these categories.
For each category, there is a list of disjoint rubric items, out of which one should be selected (i.e., you should exactly select one rubric item per category). 

\smallskip
\noindent Avoid any position biases and ensure that the order in which the responses were presented does not influence your decision. Do not allow the length of responses to influence your evaluation. Be as objective as possible. 
Format your response as a valid JSON:
\begin{verbatim}
{
    "llm_with_best_rubric": <LLM_NUMBER>,
    "llm_with_best_feedback": <LLM_NUMBER>
}    
\end{verbatim}

Use 1 to indicate the first LLM as the winner, 2 to indicate the second LLM as the winner, or 0 if there is a tie. 

\noindent \textbf{Problem Statement}: \texttt{\{problem\}}

\noindent \textbf{Grading Rubric}: \texttt{\{rubric\}}

\noindent \textbf{Instructor Solution}: \texttt{\{instructor\_soln\}}

\noindent \textbf{Student Answer}: \texttt{\{student\_answer\}}

\noindent \textbf{LLM Responses to Judge}: \texttt{\{llm\_responses\}}

\section{Benchmarking Studies}
\label{app:benchmarking}
We use OpenAI’s \othreemini in our case study based on benchmarking done by Course X at the time the course was offered (Spring 2025). Specifically, all available GPT reasoning models were tested on a sample of Course X questions. After reviewing the generated solutions, the instructional team found \othreemini’s responses to be the most satisfactory, leading us to use it as the model most closely aligned with the course’s expectations for grading and feedback. As for our prompt, recall that our pipeline uses prompts that include the problem statement, grading rubrics, instructor solution, and student answer. This decision was made based on our exploration of four different prompting strategies:
\begin{itemize}
    \item In the \naiveX prompt, we provide the LLM with only the problem statement, grading rubric, and student answer. All other strategies extend this prompt.
    \item The \fewshotX prompt includes a graded example to guide the model's evaluation.
    \item The \solutionX prompt includes the instructor's reference solution.
    \item The \chainotX prompt instructs the model to explain its reasoning process.
\end{itemize}
The final version of each strategy’s prompt was developed through 4–5 rounds of iterative refinement on proof problems not included in this benchmarking. We selected four proof problems from Course X for this purpose, two of which were \textbf{short-answer problems} and two of which were \textbf{long-answer problems}. Short-answer problems were graded using a single rubric category, \textbf{solution correctness (SC)}, which consists of multiple disjoint rubric items. Long-answer problems were graded using two rubric categories, \textbf{solution quality (SQ)} and \textbf{writing quality (WQ)}, each of which consist of multiple disjoint rubric items.

To evaluate which prompting strategy performs best, we computed three problem-level metrics that compared the \othreemini's grading with each prompting strategy to that of the assigned human grader for that problem:
\begin{itemize}
    \item \textbf{Total agreement} represents the percentage agreement between the TA-assigned rubric items and the LLM-assigned rubric items across all submissions for a given problem type. 
    \item \textbf{No-mistakes agreement} represents the percentage agreement on submissions that contain no mistakes.
    \item \textbf{Mistakes agreement} represents the percentage agreement on submissions that contain mistakes.
\end{itemize}

\begin{table}[t]
\footnotesize
    \centering
    \caption{Performance comparison of four prompting strategies in assessing our selected problems, with the results being pooled across our two problem types (objective, subjective). The subjective problem type includes two sets of results since subjective problems are graded using two rubrics (solution quality, writing quality). Each metric includes a 95\% confidence interval (shown in brackets). Results for the best performing prompting strategy across each metric and each problem type are also highlighted in bold.}
    \begin{tabular}{|c|c|c|c|c|}
    \hline
    \textbf{Problem Type} & \textbf{Strategy} & \makecell{\textbf{Total} \\ \textbf{Agreement \%}} & \makecell{\textbf{No Mistakes} \\ \textbf{Agreement \%}} & \makecell{\textbf{Mistakes} \\ \textbf{Agreement \%}} \\
       \hline
    \multirow{4}{*}{\textbf{Objective}}
        & \textbf{Naï} & 48\% [30, 70] & 80\% [60, 95] & 15\% [3, 33] \\
        & \textbf{FSh} & 45\% [30, 65] & \textbf{85\% [65, 98]} & 5\% [0, 15] \\
        & \textbf{Sol} & \textbf{55\% [43, 80]} & \textbf{85\% [70, 98]} & \textbf{25\% [15, 55]} \\
        & \textbf{CoT} & 50\% [38, 75] & \textbf{85\% [70, 98]} & 15\% [5, 38] \\
    \hline
    \multirow{4}{*}{\textbf{Subjective (SQ)}} 
        & \textbf{Naï} & 35\% [15, 55] & 48\% [25, 68] & 21\% [5, 35] \\
        & \textbf{FSh} & 13\% [0, 28] & 5\% [0, 20] & 21\% [5, 35] \\
        & \textbf{Sol} & \textbf{40\% [20, 60]} & \textbf{52\% [30, 73]} & \textbf{26\% [3, 45]} \\
        & \textbf{CoT} & 28\% [8, 45] & 33\% [15, 55] & 21\% [5, 38] \\
    \hline
    \multirow{4}{*}{\textbf{Subjective (WQ)}} 
        & \textbf{Naï} & \textbf{65\% [45, 85]} & \textbf{73\% [50, 95]} & \textbf{40\% [17, 63]} \\
        & \textbf{FSh} & 25\% [10, 45] & 23\% [5, 40] & 30\% [13, 63] \\
        & \textbf{Sol} & 60\% [35, 85] & 67\% [45, 90] & \textbf{40\% [20, 60]} \\
        & \textbf{CoT} & 50\% [30, 70] & 63\% [43, 85] & 10\% [0, 25] \\
    \hline
    \end{tabular}
    \label{tab:study1_pooled}
\end{table}

We present the results, pooled across the two problem types, in Table \ref{tab:study1_pooled}. For the short-answer problem type, we find that the \solution prompt outperforms the other prompts most often. Similarly for the long-answer problem type across both rubric types and all metrics, we observe that the \solution prompt outperforms the others most often. This led us to use the \solution prompt -- which includes the problem statement, grading rubrics, instructor solution, and student answer -- as the prompting strategy for our case study's LLM pipeline.

\end{document}